\documentclass{edm_template}
\usepackage{fancyhdr}

\fancypagestyle{specialfooter}{%
  \fancyhf{}

    \lfoot{Paper presented at the Doctoral Consortium of the 11th International Conference on Educational Data Mining, July, 2018, Buffalo, NY.}
}

\begin{document}

\thispagestyle{specialfooter}
\title{Clustering Students and Inferring Skill Set Profiles with Skill Hierarchies}
%
%
%
%

\numberofauthors{2}
\author{
%
%
\alignauthor
Alan Mishler\\
       \affaddr{Department of Statistics and Data Science}\\
       \affaddr{Carnegie Mellon University}\\
       \affaddr{Pittsburgh, PA 15213}\\
       \email{amishler@stat.cmu.edu}
\alignauthor
Rebecca Nugent\\
\affaddr{Department of Statistics and Data Science}\\
       \affaddr{Carnegie Mellon University}\\
       \affaddr{Pittsburgh, PA 15213}\\
       \email{rnugent@stat.cmu.edu}
}

\maketitle
\begin{abstract}
Cognitive diagnosis models (CDMs) are a popular tool for assessing students' mastery of sets of skills \cite{Junker2001}. Given a set of $K$ skills tested on an assessment, students are classified into one of $2^K$ latent skill set profiles that represent whether they have mastered each skill or not. Traditional approaches to estimating these profiles are computationally intensive and become infeasible on large datasets \cite{Junker2001}. Instead, proxy skill estimates can be generated from the observed responses and then clustered, and these clusters can be assigned to different  profiles \cite{Chiu2009}. 
    Building on the work of \cite{Nugent2010}, we consider how to optimally perform this clustering when not all $2^K$ profiles are possible, e.g. because of hierarchical relationships among the skills, and when not all possible profiles are present in the population. We compare hierarchical clustering and several k-means variants, including semisupervised clustering using simulated student responses. The empty k-means algorithm of \cite{Nugent2010} paired with a novel method for generating starting centers yields the best overall performance.
\end{abstract}

%


\section{Introduction} 
Cognitive diagnosis models (CDMs) provide a means of estimating which skills students have and have not mastered. Given $N$ students, $K$ skills, and $J$ items, let $\mathbf{Y}$ be an $N\times J$ binary item response matrix, where $Y_{ij}$ represents whether student $i$ got item $j$ correct or not. (While missing values can exist, they are outside the scope of this work.) CDMs also utilize a binary $J\times K$ $Q$-matrix, with each entry $q_{j,k}$ indicating whether item $j$ requires skill $k$. The $Q$-matrix is constructed by domain experts and is typically assumed to be correct.  The goal is the estimation of a latent binary vector $\mathbf{\alpha}_i = (\alpha_{i1}, \ldots \alpha_{iK})$ for $i = 1, 2, \ldots N$, representing each student's mastery of the skills.

CDMs are commonly estimated using maximum likelihood methods via the Expectation Maximization (EM) algorithm, or in a Bayesian setting using Markov Chain Monte Carlo (MCMC) simulation. However, these approaches are computationally intensive and may be infeasible given large numbers of students, items, or skills, or for particularly complex models \cite{Junker2001}.

As an alternative, \cite{Chiu2009} suggest using clustering methods such as k-means or hierarchical agglomerative clustering (HC) to group students into latent classes. The item response matrix and the $Q$-matrix are used to compute a $K$-dimensional vector for each student, representing performance on items requiring each of the $K$ skills. Those vectors are then clustered to estimate the $2^K$ profiles. Under certain conditions, \cite{Chiu2009} show that HC is a consistent procedure, correctly partitioning students with probability converging to 1. They also show that k-means performs well in simulations.

Both HC and k-means require choosing the number of clusters, and the theoretical consistency of HC requires that each of the $2^K$ profiles is sampled with probability greater than 0. Since the number of profiles is exponential in $K$, it is likely or even inevitable that not all $2^K$ profiles will be present in the sample. \cite{Nugent2010} present a variant of k-means called \textit{empty k-means} in which the number of clusters is not specified in advance and anywhere between 1 and $2^K$ clusters are returned. Their method outperforms HC and conventional k-means when fewer than $2^K$ profiles are present. 

Typically there is no way to know which profiles are present in a population, but in some cases it is possible to exclude some profiles. If skills are hierarchically arranged, then students cannot master a ``child'' skill without mastering its ``parent'' skill. For example, if addition is prerequisite to multiplication, then a profile $(0,1)$ representing mastery of multiplication without mastery of addition is not possible, but profiles $(0,0)$, $(1,0)$, and $(1,1)$ are possible. \cite{Su2013} estimates CDMs with hierarchical skill relationships by precluding profiles that are not possible under the hierarchy and then estimating model parameters using maximum marginal likelihood. Although this approach improves model fit and item parameter estimation in several simulations \cite{Su2013}, it is expected that it would become computationally intractable as the number of skills, items, and/or students grew.

We extend previous work by investigating the use of clustering methods in cases when not all profiles are present and there is information about which profiles are absent, for example because of hierarchical skill relationships. These clustering approaches are computationally scalable and are agnostic to the data-generating mechanism, making them appealing in a broad range of settings. 

\section{Method}
We use simulation to compare several clustering methods when fewer than $2^K$ profiles are present, potentially given some skill hierarchy. In this section, we describe the derivation of the capability scores, the hierarchies we examined, and the clustering methods we used.

\subsection{Sum and capability scores}
Recall the response vectors $\mathbf{Y}_i$ and $\mathbf{Q}$-matrix defined in Section 1. For clustering, \cite{Chiu2009} define, for each student, a vector $\mathbf{W}_i = (W_{i1}, W_{i2}, \ldots, W_{iK})$, where $W_{ik} = \sum_{j=1}^J Y_{ij}q_{jk}$ is the \textit{sum score} for that student for skill $k$. Each sum score represents the number of items a student got right out of items that require skill $k$. The vectors $\mathbf{W}_i$ therefore lie in a hyperrectangle touching the origin, with the length of side $k$ equal to the number of items that require skill $k$. 

Here, we follow \cite{Nugent2010} and divide each $W_{ik}$ by $k$ to produce \textit{capability scores}, which are constrained to lie in the $k$-dimensional unit hypercube. This is intuitively appealing, since the scores serve as a proxy for the latent skill profiles, which lie at the vertices of this hypercube. 

\subsection{Skill hierarchies}
\begin{figure}
	\includegraphics[width=1\linewidth]{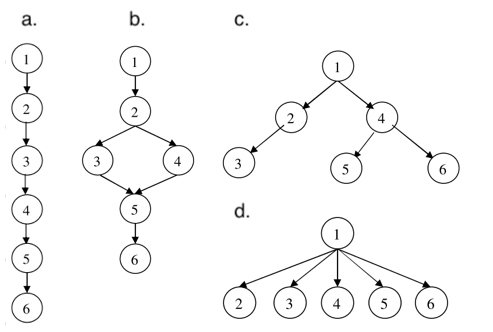}
	\caption{Four types of skill hierarchies: (a) linear, (b) convergent, (c) divergent, and (d) unstructured.}
	\label{fig:hierarchies}
\end{figure}

In the CDM literature, it is typical to assume that all $2^K$ profiles are possible, which is tantamount to assuming that skills can be mastered in any order. This assumption is often not realistic, however, as some skills may need to be mastered before others \cite{Leighton2004}. Skill hierarchies can be represented by Directed Acyclic Graphs (DAGs), with edges indicating prerequisite relationships. Although any DAG is possible in principle, we compare four common types of skill hierarchies: \textit{linear}, \textit{convergent}, \textit{divergent}, and \textit{unstructured}. Figure \ref{fig:hierarchies}, taken from Figure 1 in \cite{Leighton2004}, shows examples of these four types for a set of six skills. We also examine the case of a null hierarchy, i.e. when the skills can be learned in any order. Let $L_h$ be the number of possible possible profiles under a hierarchy $h$. With six skills, $L_\text{null} = 2^6 = 64$. The other hierarchies admit far fewer profiles: $L_\text{linear} = 7$, $L_\text{convergent} = 12$, $L_\text{divergent} = 16$, and $L_\text{unstructured} = 33$.

\subsection{Clustering methods}
\subsubsection{Hierarchical agglomerative clustering (HC)}
This method starts with each data point in its own cluster and iteratively merges the closest clusters, according to a specified distance function, until all the data points are in a single cluster. The resulting dendrogram can then be cut to produce a specified number of clusters between $1$ and $N$. We use complete linkage as the distance function and cut the dendrogram at the point representing the largest merge distance. If this produces more than $L_h$ clusters, then we instead cut the dendrogram so as to produce $L_h$ clusters.
\subsubsection{k-means}
We use traditional k-means, with the number of clusters set to $L_h$ (\cite{Hartigan1979}) via $\texttt{kmeans()}$ in $R$ with 5 random restarts. This algorithm generates an error when clusters are empty. As such, we use only random starting center observations.

\subsubsection{Empty k-means}
We use the algorithm in \cite{Nugent2010}, set to allow up to $L_h$ clusters. We select starting centers by (1) random sampling, (2) the rescaling method described in \cite{Nugent2010}, and (3) generating ``pseudocenters'' from an assumed model. The pseudocenters rely on the intuition that if the true data-generating model were known, then it could be used to select appropriate starting centers, for example by finding the expected capability score vector under each possible profile. As an approximation to this, the pseudocenters are generated by simulating data under a particular model for each possible profile and then taking the mean of each profile's capability scores. The model used to generate the pseudocenters may be different from the model used to generate the to-be-clustered data; we consider the issue of model misspecification below.

\subsubsection{Semisupervised clustering}
As with the pseudocenters, pseudodata is generated under a model for each possible profile; then semisupervised clustering is performed using the pseudodata and the real data. We use the \texttt{lcvqe()} function from the \texttt{conclust} package in R \cite{conclust2016}. The LCVQE algorithm is a k-means variant that clusters labeled and unlabeled data points while attempting to preserve the clustering implied by the labels. This technique initially produces $L_h$ clusters, but some of those clusters may end up containing only the labeled data, in which case they are discarded and a smaller number of clusters of the (originally) unlabeled data remain.

\section{Simulation Results}
\subsection{Framework}
The numbers of skills $K$, items $J$, and students $N$ were 6, 30, and 250, respectively. We sampled a single Q matrix of 9 single-skill items (30\%), 18 two-skill items (60\%), and 3 three-skill items (10\%). For each skill hierarchy $h$, subsets of profiles of size $3, 4, \ldots, L_h$ were selected. For the linear, convergent, and divergent hierarchies, the subsets consisted of rows 1-3, 1-4, $\ldots$, 1-$L_h$ from the matrices in \cite{Su2013} representing the possible profiles. For the unstructured and null hierarchies, the subsets were sampled at random.

Item responses were generated from two CDMs, a DINA and a NIDA \cite{Junker2001}. The item response functions are:
\begin{align*}
	P(y_{ij} = 1|\eta_{ij}, s_j, g_j) &= (1 - s_j)^{\eta_{ij}}g_j^{1-\eta_{ij}} \tag{DINA} \\
    P(y_{ij} = 1|\alpha_i, s_k, g_k) &= \prod_{k=1}^K[(1-s_k)^{\alpha_{ik}}g_k^{1-\alpha_{ik}}]^{q_{jk}}  \tag{NIDA}
\end{align*}
where $\alpha_i$ is the skill set profile, $\eta_{ij} = \prod_{k=1}^K \alpha_{ik}^{q_{jk}}$ indicates whether student $i$ has mastered all the skills necessary for item $j$, and $s_j, s_k, g_j, g_k$ are slip and guess parameters indexed on items and skills, respectively. The slip parameters represent the probability of responding incorrectly even with sufficient skills; the guess parameters represent the probability of guessing correctly without those skills. The slip and guess parameters were sampled from a Uniform(0, 0.30) and a Uniform(0, 0.15) distribution, respectively. Item responses were converted to capability scores for clustering. Clustering performance was assessed via the Adjusted Rand Index (ARI), a measure of agreement between two partitions \cite{Hubert1985}. (Here, the second partition consisted of the true profile labels). The expected ARI for a pair of random partitions is 0, while identical partitions yield the maximum ARI of 1.

\subsection{Results}
\begin{figure*}
	\includegraphics{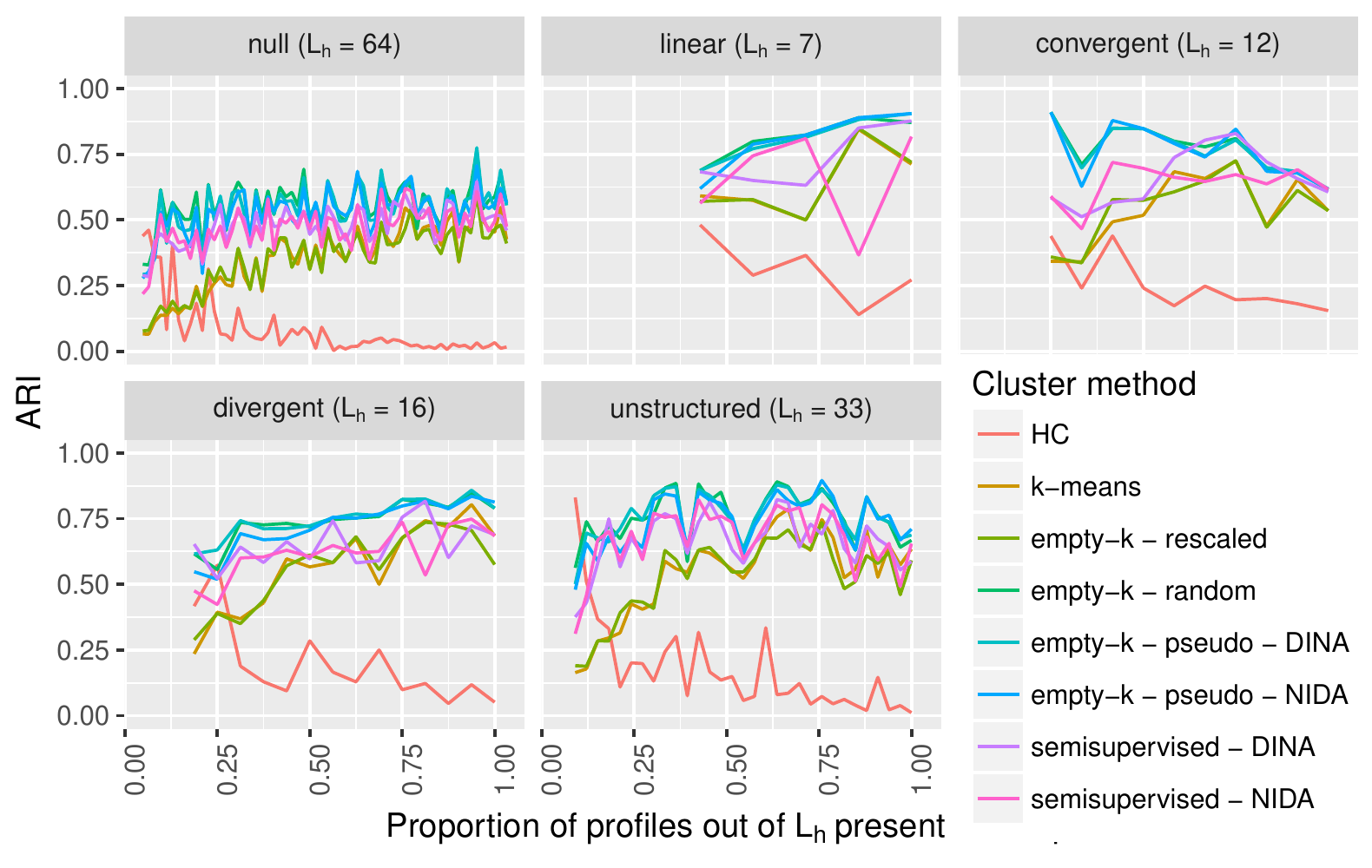}
	\caption{ARIs for five hierarchies and four clustering methods, for DINA-generated data}
	\label{fig:results_dina}
\end{figure*}

\begin{figure*}
	\includegraphics{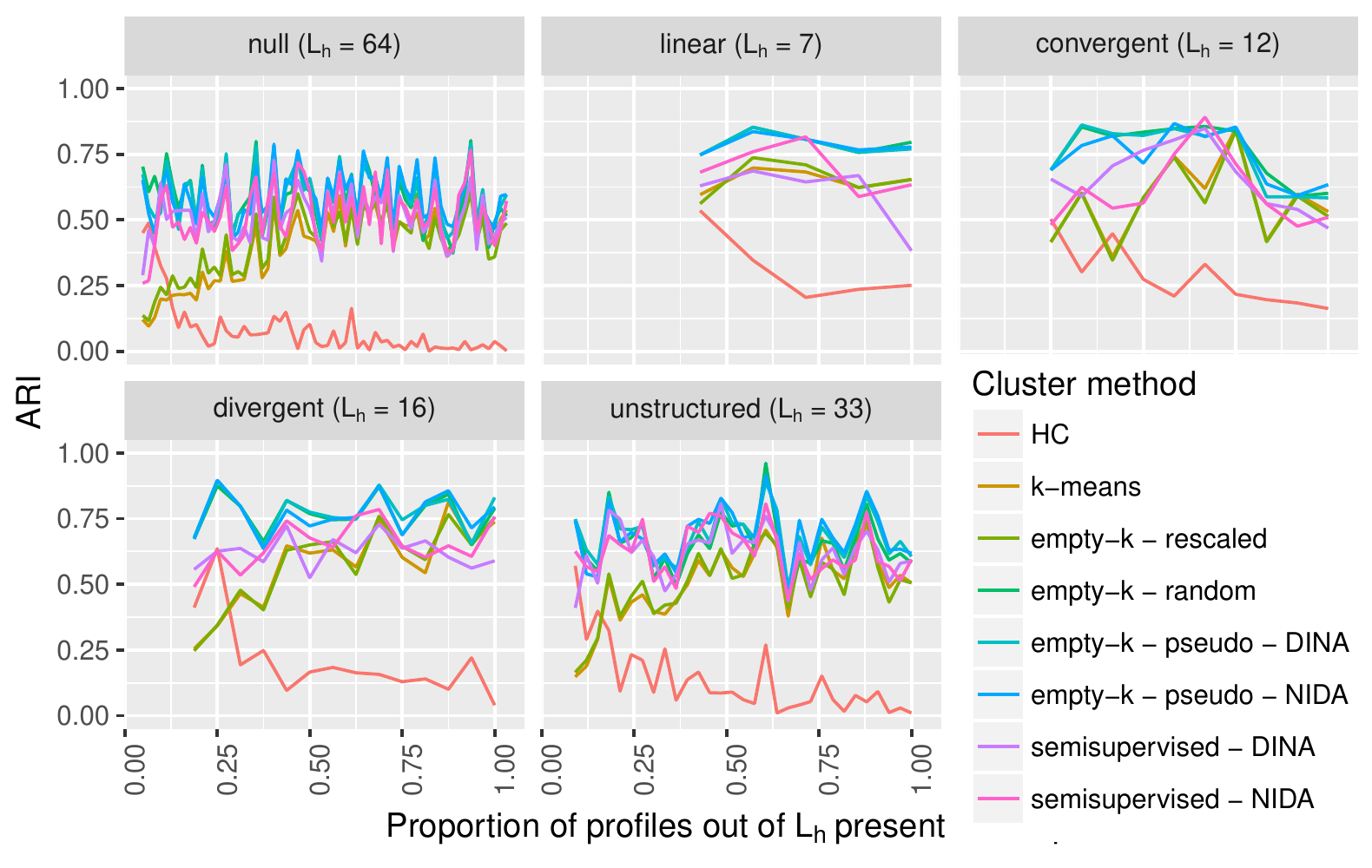}
	\caption{ARIs for five hierarchies and four clustering methods, for NIDA-generated data}
	\label{fig:results_nida}
\end{figure*}

Figures \ref{fig:results_dina} and \ref{fig:results_nida} shows ARIs for data generated from the DINA and NIDA, respectively. The x-axes show the proportion of profiles present out of the number $L_h$ that are possible under each hierarchy. The four ``emptyK'' methods refer to empty k-means with four different sets of starting centers. The ``DINA'' and ``NIDA'' suffixes refer to the models used to generate pseudodata; these models were independent of the models used to generate the ``real'' data.

The DINA and NIDA data show similar results. Empty k-means with pseudocenters performed the best in nearly every case. Surprisingly, this method performed well even under model misspecification, e.g. when a NIDA was used with an underlying DINA. The hierarchical clustering method performed moderately well with small numbers of profiles, but performance dropped off quickly as that number increased. These patterns held even though both $k$-means and HC were constrained to a maximum of $L_h$ clusters, a more favorable scenario than the traditional $2^K$ clusters.

Performance did not increase consistently in the proportion of profiles, even though the clustering methods were all set to allow up to $L_h$ clusters, which might be expected to lead to too many clusters for small profile subsets. This may reflect the fact that more profiles make the score space more crowded, which makes it difficult to distinguish clusters.

The ARI fluctuations in the null and unstructured hierarchies are likely due to the particular profiles that were sampled. The methods appear to be closely correlated in these cases, suggesting that some subsets are harder to cluster than others, probably due to skill estimate attenuation \cite{Nugent2010}.

Semisupervised clustering performed moderately well, but again, the correctly specified model did not necessarily outperform the misspecified model.

\section{Discussion and Future Work}
Previous work has suggested using k-means \cite{Chiu2009} or empty k-means \cite{Nugent2010} to partition students into classes based on their latent skill set profiles, while \cite{Su2013} used a likelihood-based approach to estimate CDMs in the presence of skill hierarchies. We synthesized and extended these lines of research, applying a range of clustering approaches when skill hierarchies render some profiles impossible and when not all possible profiles are present in a sample. A novel approach that combines the empty k-means algorithm with starting centers generated from pseudodata yielded the best performance, even when the generating model was misspecified.

In future work, we intend to investigate further how robust the empty k-means method with pseudocenters is to model misspecification. Both the DINA and the NIDA are conjunctive CDMs, so it is possible that performance would degrade if they were paired with disjunctive models. We plan to investigate the fluctuations in the null and unstructured hierarchy results to understand what kinds of profile subsets are easiest to distinguish. Finally, we intend to explore allowing for misspecification of the skill hierarchy, for example by translating it into a set of soft rather than hard constraints; and we will consider ways to infer the skill hierarchy when it is unknown or only partially known. 

The pseudodata-based approaches have an advantage in that they yield not just partitions but profile labels, since the final clusters are associated with existing, labeled data points. With the other methods, labels must be derived by, for example, assigning clusters to the nearest licit vertex of the capability score space. We think this is a promising approach that merits further investigation.


%
\bibliographystyle{abbrv}
\bibliography{references}  

\begin{thebibliography}{1}

\bibitem{Chiu2009}
C.-Y. Chiu and J.~Douglas.
\newblock {Cluster analysis for cognitive diagnosis: Theory and applications}.
\newblock {\em Psychometrika}, 74(4):633--665, 2009.

\bibitem{Hartigan1979}
J.~A. Hartigan and M.~A. Wong.
\newblock {A K-Means Clustering Algorithm}.
\newblock {\em Applied Statistics}, 28(1):100--108, 1979.

\bibitem{conclust2016}
T.~K. Hiep and N.~M. Duc.
\newblock {\em conclust: Pairwise Constraints Clustering}, 2016.
\newblock R package version 1.1.

\bibitem{Hubert1985}
L.~Hubert and P.~Arabie.
\newblock {Comparing partitions}.
\newblock {\em Journal of Classification}, 2(1):193--218, 1985.

\bibitem{Junker2001}
B.~W. Junker and K.~Sijtsma.
\newblock {Cognitive Assessment Models with Few Assumptions, and Connections
  with Nonparametric Item Response Theory}.
\newblock {\em Applied Psychological Measurement}, 25(3):258--272, 2001.

\bibitem{Leighton2004}
J.~P. Leighton, M.~J. Gierl, and S.~M. Hunka.
\newblock {The attribute hierarchy method for cognitive assessment: A variation
  on Tatsuoka's rule-space approach}.
\newblock {\em Journal of Educational Measurement}, 41(3):205--237, 2004.

\bibitem{Nugent2010}
R.~Nugent, N.~Dean, and E.~Ayers.
\newblock {Skill set profile clustering: the empty K-means algorithm with
  automatic specification of starting cluster centers}.
\newblock {\em Proceedings of the 3rd International Conference on Educational
  Data Mining}, pages 151--160, 2010.

\bibitem{Su2013}
Y.-L. Su.
\newblock {Cognitive diagnostic analysis using hierarchically structured
  skills}.
\newblock {\em ProQuest Dissertations and Theses}, 2013.

\end{thebibliography}
%
%

\balancecolumns
\end{document}